\documentclass{elsart}

 \usepackage{graphicx}

\usepackage{amssymb}

\begin{document}

\begin{frontmatter}



\title{Equal energy sharing double photo ionization of the Helium
atom at 20eV and 40eV above threshold}


\author[cu]{J. N. Das}
\author[scc]{K. Chakrabarti \corauthref{cor}}
\corauth[cor]{Corresponding Author} \ead{kkch@eth.net}
\author[cu]{S. Paul}
\address[cu]{Department of Applied Mathematics, University College
of Science, 92 Acharya Prafulla Chandra Road, Calcutta - 700 009,
India.}
\address[scc]{Department of Mathematics, Scottish Church College, 1 \& 3
Urquhart Square, Calcutta - 700 006, India}

\begin{abstract}
    In this article we present triple differential cross sections
for equal energy sharing kinematics for double photoionization of
the helium atom at 20 and 40eV above threshold in the framework of
the hyperspherical partial wave theory. This supplements our
earlier work \cite{DCP03} in which we were successful in showing
fully, gauge independence of the results in our formalism. Also in
this work we treat cases in which the Stokes parameter S$_1 < 1$
so that partial polarization of the photon source is also taken
into account. Agreement in shape with the Convergent Close
Coupling \cite{BK03,HB98} calculation and the experiments appears
to be excellent.
\end{abstract}

\end{frontmatter}

\section{Introduction}
\label{sec1}   Double photo ionization (DPI) of the helium is one
of those fundamental atomic processes which has received wide
attention from both theorists and experimentalists. Helium is one
of the most abundant elements in stellar atmospheres and hence
analysis of Helium spectra is of considerable importance in
astrophysics. From the theoretical aspect, the absorption of a
photon and subsequent ejection of both electrons from the Helium
atom into the continuum, leads to a pure Coulomb three body
problem which, at low incident energies, involve complex short
range interactions and strong electron correlations  that are
difficult to treat theoretically and hence present a challenge to
theorists.

    Detailed discussion about existing theories of helium
double photo ionization is presented in our earlier work
\cite{DCP03}. It would still be worthwhile to mention the
convergent close coupling (CCC) theory of Bray \emph{et al}
\cite{KB98}, 2SC theory of Pont and Shakeshaft \cite{PS95a} and
the 3C theory of Maulbetsch and Briggs \cite{MB93} since all of
them have been very widely used to study the double photo
ionization process. It would also be very appropriate to single
out and mention the hyperspherical $\mathcal{R}$-matrix with semi
classical outgoing waves (H$\mathcal{R}$M-SOW) theory of Malegat
\emph{et al} \cite{MSL00,SML02} for two reasons. First, the method
has been quite successful in describing the total ionization cross
section (TICS), single and triple differential cross sections
(SDCS and TDCS). Secondly the method bears similarity to our
calculation in the use of hyperspherical coordinates, though the
techniques used in extracting scattering information is very
different from ours. The H$\mathcal{R}$M-SOW method uses
$\mathcal{R}$-matrix \cite{RB75} method to obtain solution of the
radial wave equation in the inner region bounded by a hypersphere
at $R_0$ ($R$ being the hyperradius). Beyond $R_0$ the solution of
the radial equations are propagated semiclasically (see section IV
of \cite{SML02}). In contrast our method is fully quantum
mechanical and free from any approximations, though perhaps it is
more demanding on the computational resources.

    The helium double photo ionization is similar to the electron
impact ionization of atomic hydrogen. However, due to the
$^1$S$^e$ symmetry of the initial state, the final state can be
only of $^1$P$^o$ symmetry. This leads to considerable
simplification in computations in contrast to e-H ionization, in
which there are many more contributing states in the final
channel. Also there are selection rules that are valid only for
double photo ionization \cite{MB93} (for a recent review see
\cite{BS00}).

    As is well known, the DPI triple differential cross-section
can be calculated in three gauges, namely length, velocity and
acceleration gauges. In principle the TDCS must be independent of
the choice of gauge. However, it has been shown by Lucy {\it{et
al}} \cite{LRW} that the TDCS results are notoriously gauge
dependent unless one uses very accurate wave functions for both
initial and final channels. In our previous work \cite{DCP03} we
had been able to show that our results are largely gauge
independent and length and velocity gauge results obtained were
nearly identical. We now have the TDCS results in the acceleration
gauge also, and we will make a brief comparison between the TDCS
in the three gauges below before we study other cases. Also in our
work \cite{DCP03} we obtained TDCS at 20eV incident photon energy
and our results were for incident photon source linearly polarized
with Stokes parameter $S_1=1$. Having obtained an essentially
gauge independent formalism, in this work we present double photo
ionization TDCS at 20eV and 40eV above the helium double
ionization threshold 79eV. Also we choose situations in which the
incident photon beam is partially \linebreak (linear) polarized,
the degree of polarization being given by the Stokes \linebreak
parameter S$_1$ (for a definition of Stokes parameters see
\cite{BS00}. Our $S_1$ corresponds to $S_{lin}$ given in section
3.2 of Ref. \cite{BS00} in which the $x$ and $y$ axes are
respectively along the major and minor axes of the polarization
ellipse).

        The double ionization TDCS is obtained from the transition
matrix elements given by
\begin{equation}
T_{fi} = \langle \Psi_f^{(-)}|V_i|\Phi_i \rangle,
\end{equation}
where $\Phi_i(\pol{r}_1,\pol{r}_2)$ is the helium ground state,
$\pol{r}_1$ and  $\pol{r}_2$ being the coordinates of the two
electrons with respect to the helium nucleus (assumed to be at
rest). $\Psi_f^{(-)}$ is the final channel continuum state and
$V_i$ is the interaction term given by
\begin{equation}
V_i = \pol{\epsilon}\cdot \pol{D}.
\end{equation}
In (2) $\pol{D}$ is the dipole operator given by $\pol{D} =
\pol{\nabla}_1 + \pol{\nabla}_2$ (velocity form), \linebreak
$\pol{D} = \omega_i(\pol{r}_1 + \pol{r}_2)$ (length form) or
$\pol{D} = -\frac{1}{\omega_i}(\pol{\nabla}_1 + \pol{\nabla}_2) V$
(acceleration form), where
\begin{displaymath}
V = -\frac{1}{r_1} - \frac{1}{r_2} + \frac{1}{|\pol{r}_1 -
\pol{r}_2|}
\end{displaymath}
is the full three body interaction potential and $\pol{\epsilon}$
is the photon polarization vector. For an arbitrary degree of
(linear) polarization of the incident photon beam characterized by
the Stokes parameter S$_1$, the TDCS is given by (see \cite{CWR00}
for example and references therein)
\begin{eqnarray}
\frac{d^3\sigma}{d \Omega_1 d \Omega_2 dE_1} = (\sigma_x +
\sigma_y) + \frac{S_1}{2}(\sigma_x - \sigma_y),
\end{eqnarray}
where $\sigma_x$ and $\sigma_y$ refer to TDCS calculated with
polarization vectors along the x-axis and y-axis respectively.

    A schematic diagram of the scattering geometry is shown in
figure 1. The photon emerges from the bottom of the scattering
plane along the z-axis. The two photo-electrons are ejected after
collision in the direction $\theta_1$ and $\theta_2$ in the
scattering plane.

\begin{figure}[h]
\begin{center}
\includegraphics[bb=120 340 493 559,scale=0.5]{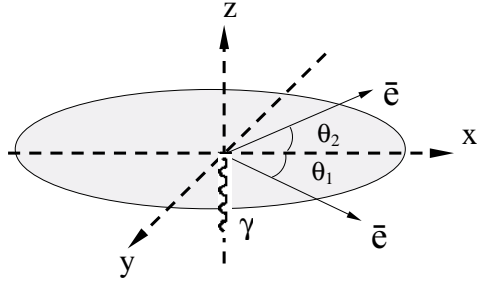}
\end{center}
\caption{ Schematic diagram of the scattering geometry.}
\end{figure}

\section{Theory}
\label{sec2}
    The hyperspherical partial wave theory is discussed in
considerable detail in our earlier work \cite{DCP03}. Therefore we
give here only the essentials.

    For the Helium ground state we take a highly correlated 20 term
Hylleraas type wave function given by Hart and Herzberg \cite{HH}.
To calculate the final channel continuum state $\Psi_f^{(-)}$ we
use hyperspherical coordinates $R = \sqrt{r_1^2 + r_2^2}$, $\alpha
= \arctan(r_2/r_1)$, $\hat{r}_1 = (\theta_1, \phi_1)$, $\hat{r}_2
= (\theta_2, \phi_2)$ and $\omega = (\alpha, \hat{r}_1,
\hat{r}_2)$. Also we set $P = \sqrt{p_1^2 + p_2^2}$, $\alpha_0 =
\arctan(p_2/p_1)$, $\hat{p}_1 = (\theta_{p_1}, \phi_{p_1})$,
$\hat{p}_2 = (\theta_{p_2}, \phi_{p_2})$ and $\omega_0 =
(\alpha_0, \hat{p}_1, \hat{p}_2)$ where $\pol{r}_i$ and
$\pol{p}_i$ (i = 1, 2) are the coordinates and momenta of the
$i^{th}$ particle. $\Psi_f^{(-)}$ is then expanded in
hyperspherical harmonics  \cite{DAS98,LIN} that are functions of
the five angular variables and $\ell_1$, $\ell_2$, $n$, $L$, $M$,
which are respectively the angular momenta of the two electrons,
the order of the Jacobi polynomial in the hyperspherical
harmonics, the total angular momentum and its projection. For a
given symmetry $s$ we decompose the final state as
\begin{equation}
\Psi_{fs}^{(-)}(R, \omega)=\sqrt{\frac{2}{\pi}} \sum_{\lambda}
\frac{F_{\lambda}^s(\rho)}{{\rho}^{\frac{5}{2}}}\;
\phi_{\lambda}^s(\omega)
\end{equation}
where  $\lambda$ is the composite index ($\ell_1$, $\ell_2$, $n$,
L, M) or $2n + \ell_1 + \ell_2$ depending  on the context and
$\rho = PR$ and $\phi_{\lambda}^s(\omega)$ are are orthogonal
functions that are a product of a Jacobi polynomial $P_n^{l_1
l2}(\alpha)$ and a coupled angular momentum eigenfunction
$\mathcal{Y}_{l_1 l_2}^{LM}(\Omega_1, \Omega_2)$.

    $F_\lambda^s$ then satisfy the infinite set of coupled
differential equations
\begin{equation}
\Big[ \frac{d^2}{d\rho^2} +1 -
\frac{\nu_{\lambda}\,(\nu_{\lambda}+1)} {\rho^2}\Big]
F_{\lambda}^s (\rho) + \sum_{\lambda'} \frac{2\;\alpha_{\lambda
\lambda'}^s}{P\rho} \ F_{\lambda'}^s(\rho) = 0.
\end{equation}
Here $\alpha_{\lambda \lambda'}^s$ are the matrix elements of the
full three-body  interaction potential and $\nu_{\lambda}=\lambda
+\frac{3}{2}$. Since the final channel state must have the
$^1$P$^o$ symmetry $s$ is fixed. The contributing radial waves
then correspond to $L=1$ and (odd) parity $\pi= -1$ so that
writing $N = (\ell_1, \ell_2, n)$ and $F_{\lambda}^s = f_N$ the
equations for the relevant set of radial waves become
\begin{equation}
\Big[ \frac{d^2}{d\rho^2} +1 - \frac{\nu_N\,(\nu_N+1)}
{\rho^2}\Big]f_N + \sum_{N'} \frac{2\; \alpha_{NN'}} {P\rho} \,
f_{N'} = 0.
\end{equation}
For actual computations we truncate the set of equations to some
maximum value $N_{mx}$ of $N$. These $N_{mx}$ equations in
$N_{mx}$ variables are solved from origin to infinity.
Construction of the radial wave solution is presented in our
earlier works \cite{DCP03,DPC02} with considerable rigor. So we
omit the details in this work. Knowing the radial wave solution,
the final channel state can be found from (4) and the transition
matrix elements from (1). The photoionization TDCS can then be
obtained using
\begin{equation}
\frac{d^3\sigma}{d\Omega_1 d\Omega_2 dE_1} = \frac{2\pi^2\alpha
p_1p_2}{\omega_i} |T_{fi}|^2.
\end{equation}

\section{Results and discussion}
\label{sec3}

    Convergence in our cross sections depend on two parameters,
namely the number of coupled channels included in the computations
and the asymptotic range parameter $R_\infty$ where the asymptotic
and the interior solutions are matched (see \cite{DCP03} and
\cite{DPC02} for details). Numerical investigations show that
convergence in TDCS is obtained at $R_\infty$ = 30000 a.u. with 90
coupled channels. On the contrary oscillations in the SDCS persist
even on increasing the asymptotic range parameter $R_\infty$
substantially beyond 30000 a.u. This is in contrast to the
H$\mathcal{R}$M-SOW method where oscillations in the SDCS die out
on extending the asymptotic range. However, our computed SDCS at
$E/2$ ($E$ being the excess energy) remains more or less constant
with the increase of convergence parameters. This strongly
suggests that we employ a scaling technique similar to that used
in the CCC theory (see [17]). Whenever the true SDCS at $E/2$ is
known (from experiment or from some other theory, for example
\cite{PS95,KB00}) we normalize our computed SDCS at $E/2$ to the
true SDCS by multiplying with a factor. Subsequently, we use this
factor to scale our TDCS. Thus the TDCS at 20 eV incident energy
have been scaled with a factor 0.8 and at 40 eV by a factor 0.6.
In this way we are able to obtain absolute TDCS.

\subsection{Gauge independence}
    As remarked earlier, in our previous work \cite{DCP03} we had
been able to show that our work is largely gauge independent.
There, independence of the TDCS results with respect to the length
and velocity gauges had been presented. We now present the TDCS
results in the acceleration gauge also. These are displayed in
figure 2. Except near the peaks, the length and velocity gauge
results are identical. The maximal departure of the length gauge
results occur near the peaks, and are respectively 20\% at
$\theta_1=0^o$, 14\% at $\theta_1=30^o$, 14\% at $\theta_1=60^o$
and 7\% at $\theta_1=90^o$ from the velocity gauge. The
acceleration gauge results are indistinguishable from that of the
velocity gauge. The gauge independence in our results is a strong
signature that our final channel wave function has the correct
asymptotic and short range behaviour. Since the velocity and
acceleration gauge results are identical, in subsequent
calculations we compute TDCS in the acceleration gauge only, as
this makes our computations much simpler for values of the Stokes
parameter $S_1 < 1$.

\begin{figure}[h]
\begin{center}
\includegraphics[width=14cm]{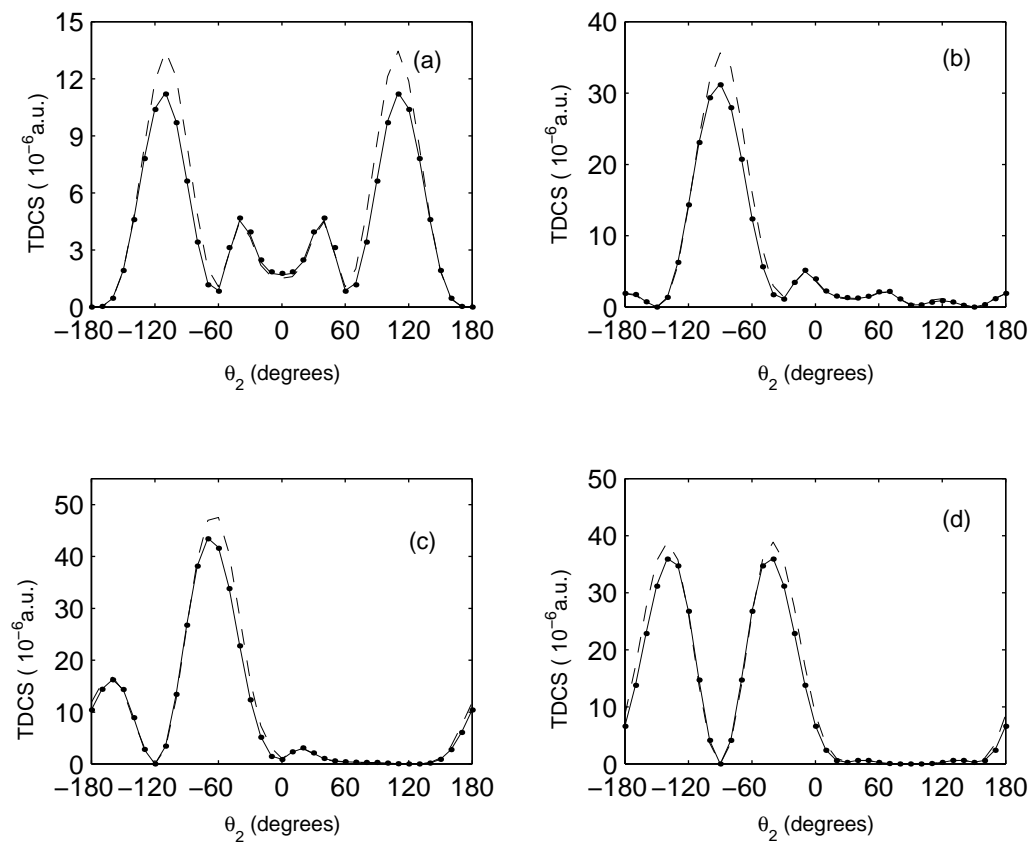}
\end{center}
\caption{TDCS for equal energy sharing double photo ionization of
the helium atom at 99eV incident photon energy in units of
$10^{-6}$ a.u. and (a) $\theta_1 = 0^o$, (b) $\theta_1 = 30^o$,
(c) $\theta_1 = 60^o$, (d) $\theta_1 = 90^o$. Results are for
length gauge (dashed curve), velocity gauge (continuous curve) and
acceleration gauge (dotted curve). The Stokes parameter $S_1 =
1.0$. }
\end{figure}

\subsection{Results for 20eV excess energy}
In this section we display our results for 99eV incident photon
energy and include polarization states by selecting cases with
Stokes parameter $S_1 < 1$.

\begin{figure}[h]
\begin{center}
\includegraphics[width=7cm]{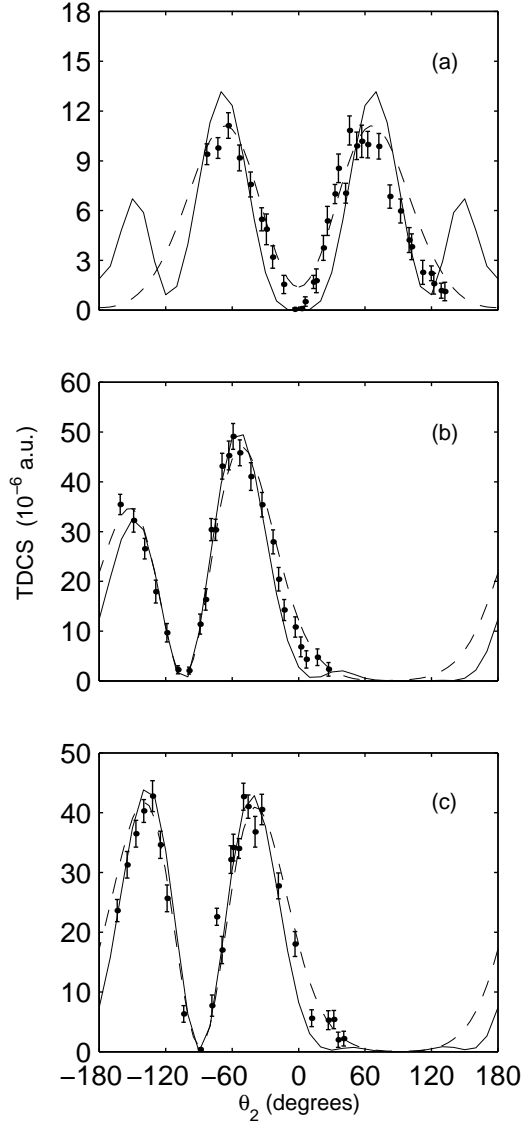}
\end{center}
\caption{TDCS for equal energy sharing double photo ionization of
the helium atom at 99eV incident energy in units of $10^{-6}$ a.u.
for (a) $\theta_1 = 180^o,\; S_1 =0.57$, (b) $\theta_1 = -76^o,\;
S_1 =0.56$, (c) $\theta_1 = -91^o,\; S_1 =0.53$. Theory:
continuous curve - present results; dashed curve - CCC results
from \cite{BK03}. In (b) and (c) the CCC results have been scaled
with a factor 0.52. Experiment: Filled circles with errorbars are
from \cite{BS00} normalized suitably with the present results in
each figure.}
\end{figure}

    In figure 3 we compare our results with experimental data from
Briggs and Schmidt \cite{BS00} and the results of the CCC theory
\cite{BK03,HB98} which are on an absolute scale. Since the
experimental data are not absolute we have normalized the data in
each set to our computed results. Also the CCC results in figure 3
(b) and (c) have been scaled with a factor 0.52. The agreement of
the present results with the experiment and the CCC results
appears to be good everywhere in shape, except for secondary peaks
near $\theta_2 = \pm 150^o$ at $\theta_1 = 180^o$ (figure 3 (a)).
To see whether this additional structure in figure 3 (a) remains,
we may need to perform a much larger calculation, that involves
the inclusion of many more partial waves in our final state wave
function.

\begin{figure}[h]
\begin{center}
\includegraphics[width=7cm]{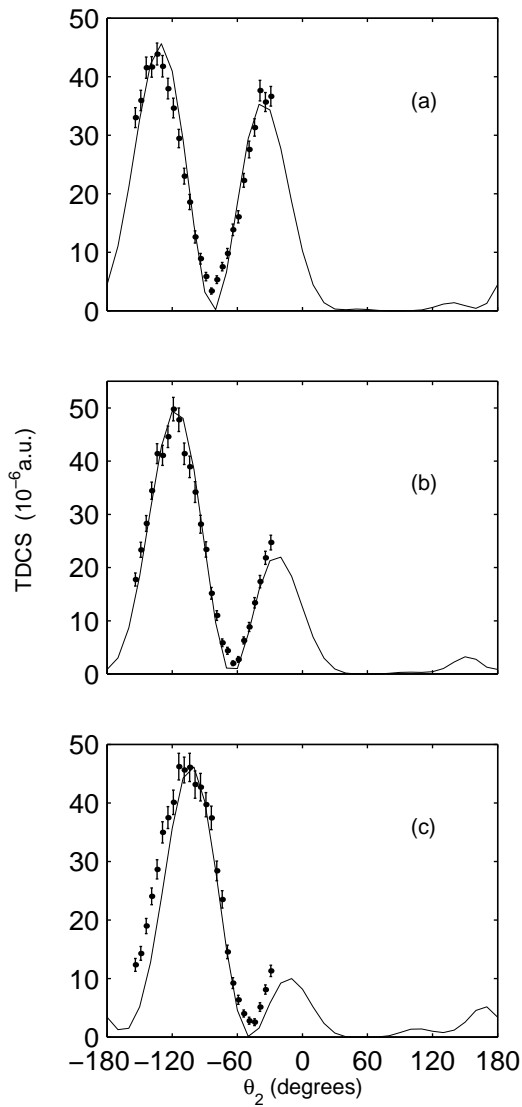}
\end{center}
\caption{TDCS for equal energy sharing double photo ionization of
the helium atom at 99eV incident energy in units of $10^{-6}$ a.u.
Theory: (a) $\theta_1 = 98^o$, (b) $\theta_1 = 115^o$, (c)
$\theta_1 = 132^o$ The Stokes parameter $S_1 = 0.67$. Experiment:
Filled circles with errorbars are from \cite{WCR98} normalized
suitably with the present results by multiplying with a
\emph{single factor}.}
\end{figure}

 In figure 4 we compare our results with the experimental
data of Weightman \emph{et al} \cite{WCR98}. The experimental data
is not absolute and have been normalized to the present results by
multiplying all data with a single factor. The agreement is
excellent everywhere.

\begin{figure}[h]
\begin{center}
\includegraphics[bb=64 204 459 606,scale=0.9]{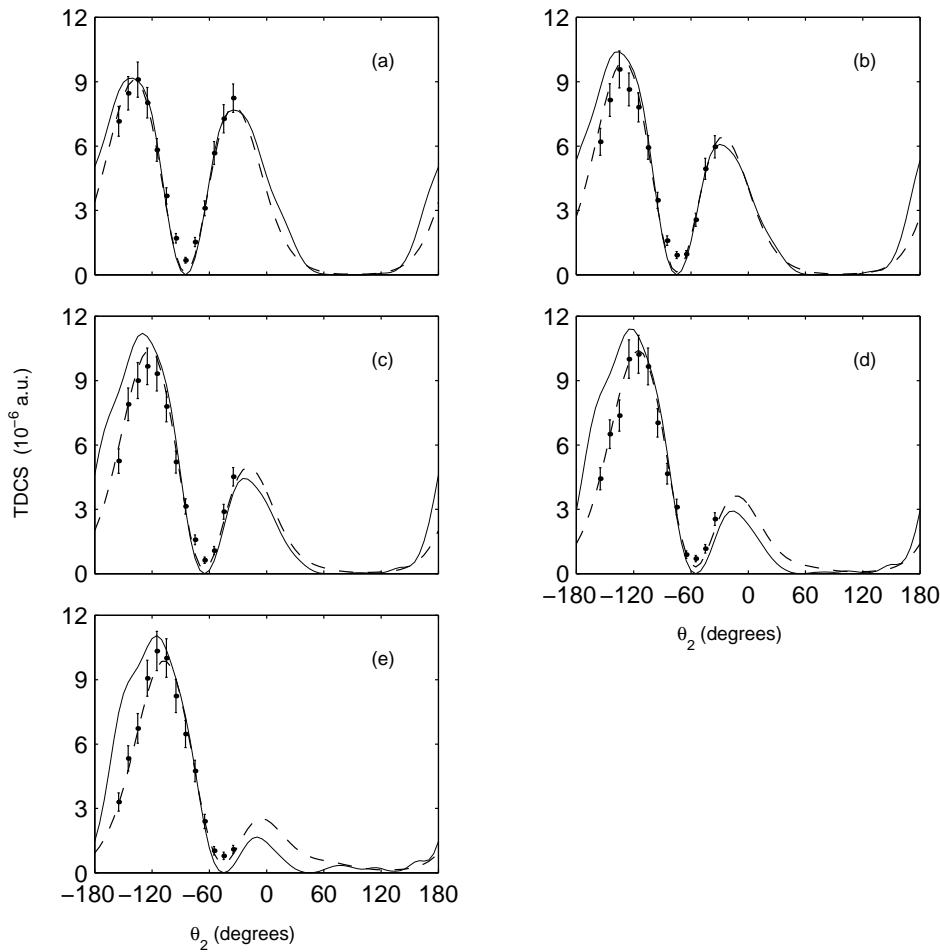}
\end{center}
\caption{TDCS for equal energy sharing double photo ionization of
the helium atom at 119eV incident energy in units of $10^{-6}$.
Theory: (a) $\theta_1 = 95^o$, (b) $\theta_1 = 105^o$, (c)
$\theta_1 = 115^o$, (d) $\theta_1 = 125^o$, (e) $\theta_1 =
135^o$. The Stokes parameter $S_1 = 0.8$. Experiment: Filled
circles with errorbars are from \cite{CWR00} normalized suitably
with the present results by multiplying with a \emph{single
factor}.  The CCC results have been scaled by 0.4 for all
figures.}
\end{figure}

There is the slight hint of a peak in figure 4 (a) near $\theta_2
= 180^o$ and its magnitude appears to increase in figures 4 (b)
and (c). Again, confirmation of this or otherwise have to be
deferred due to present limitations in our computational
resources. The CCC results for this set of data are not available
and hence comparisons are not made.

\subsection{Results for 40eV excess energy}

    Figure 5 shows our results for excess energy photon energy $E=40$
eV and Stokes parameter $S_1 = 0.8$. Shown also are the
experimental results from \linebreak Cvejanovi\'c {\it{et al}}
\cite{CWR00} and the CCC results presented in Ref. \cite{CWR00}.
The experimental points have been scaled to our theory by
multiplying with 0.05 in all cases. The absolute CCC results have
also been scaled by the same factor 0.4. Agreement in shape with
the experimental results appear to be excellent everywhere, except
in figures 5 (d) and (e) where there are slight departures around
$\theta_2=150^0$. The CCC results are also almost identical with
respect to the shapes.

\section{Conclusions}

    In this work, we have presented results for equal energy
sharing double photo ionization of the helium atom at 20 eV and 40
eV excess energy. Gauge independence of our TDCS results are
shown. Cases in which the incident photon beam is partially
polarized are considered. Comparisons are made with the
experiments and the CCC theory wherever available and the results
are seen to be consistent in shape. In the absence of absolute
TDCS measurements for the chosen kinematics, it is difficult to
say anything about the correctness of the magnitude of the various
results. We also mention that our results presented in this work
have converged approximately. For getting fully converged results,
more computational facilities may be necessary. All the
computations reported here were done on desktop computers with
Pentium IV class CPU and 512M core memory.

    In a future work, we propose to deal with unequal energy
sharing kinematics. However, due to computational limitations we
cannot reproduce results for extremely asymmetric energy sharing
at this moment. As noted in our work \cite{DPC02} high Rydberg
states tend to interfere with our continuum state giving
undesirable results in such cases. To cope with these situations,
considerably more computational resources may be necessary.

\section{Acknowledgements}

    The authors are grateful to V. Schmidt and T. J. Reddish for
providing the experimental results and to Igor Bray and Anatoli
Kheifets for providing the CCC results electronically. KC
acknowledges support from the \linebreak University Grants
Commission in the form of a Minor Research Project \linebreak
F.PSW-035/02(ERO). SP is grateful to CSIR for providing a research
\linebreak fellowship.



\begin{thebibliography}{30}

\bibitem{DCP03}
J.N. Das, K. Chakrabarti and S. Paul, J. Phys. B: At. Mol. Opt.
Phys. 36(2003)2707.

\bibitem{KB98}
A. S. Kheifets and I. Bray J. Phys. B: At. Mol. Opt. Phys.
31(1998) L447

\bibitem{PS95a}
M. Pont and R. Shakeshaft Phys. Rev. A51(1995)R2676.

\bibitem{MB93}
F. Maulbetsch, J. S. Briggs, J. Phys. B: At. Mol. Opt. Phys. 26
(1993)1679

\bibitem{MSL00}
L. Malegat, P. Selles, A. K. Kazansky, Phys. Rev. Lett.
85(2000)4450.

\bibitem{SML02}
P. Selles, L. Malegat, A. K. Kazansky, Phys. Rev. A65(2002)032711.

\bibitem{RB75}
P. G. Burke, W. D. Robb., Adv. At. Mol. Phys. 11(1975)143.

\bibitem{BS00}
J. S. Briggs, V. Schmidt, J. Phys. B: At. Mol. Opt. Phys. 33
(2000)R1.

\bibitem{LRW}
S. P. Lucy, J. Rasch, C. T. Whelan, H. R. J. Walters, J. Phys. B:
At. Mol. Opt. Phys. 31(1998)1237.

\bibitem{HH}
J. F. Hart, G. Herzburg, Phys. Rev. 112(1957)79.

\bibitem{DAS98}
J. N. Das, Pramana J. Phys. 50(1998)53.

\bibitem{LIN}
C. D. Lin, Phys. Rev. A 10(1974)1986.

\bibitem{DPC02}
J. N. Das, S. Paul, K. Chakrabarti, Phys. Rev. A67(2003)042717.

\bibitem{BK03}
A. S. Kheifets (2003) private communications.

\bibitem{PS95}
M. Pont, R. Shakeshaft, J. Phys. B: At. Mol. Opt. Phys.
28(1995)L571.

\bibitem{KB00}
A. S. Kheifets, I. Bray, Phys. Rev. A 62(2000)065402.

\bibitem{HB98}
H. Br\"auning, R. D\"orner, C. L. Cocke, M. H. Prior, B.
Kr\"assig, A. S. Kheifets, I. Bray, A. Br\"auning-Demian, K.
Carnes, S. Dreuil, V. Mergel, P. Richard, J. Ulrich and H.
Schmidt-B\"ocking, J. Phys. B: At. Mol. Opt. Phys. 31(1998)5149.

\bibitem{WCR98}
J. P. Wightman, S. Cvejanovi\'c, T. J. Reddish, J. Phys. B: At.
Mol. Opt. Phys. 31(1998)1753.

\bibitem{CWR00}
S. Cvejanovi\'c, J. P. Wightman, T. J. Reddish, F. Maulbetsch, M.
A. MacDonald, A. S. Kheifets, I. Bray, J. Phys. B: At. Mol. Opt.
Phys. 33(2000)265.




\end{thebibliography}
\end{document}